\documentclass[twocolumn,showpacs,preprintnumbers,amsmath,amssymb]{revtex4}
\usepackage{graphicx}
\usepackage{dcolumn}
\usepackage{bm}

\def\ube13{UBe$\rm_{13}$}

\def\bi2212{Bi$\rm_2$Sr$\rm_2$CaCu$\rm_2$O$\rm_8$}
\def\ybi2212{Bi$\rm_2$Sr$\rm_2$YCu$\rm_2$O$\rm_8$}
\def\ycabi2212{Bi$\rm_2$Sr$\rm_2$Ca$\rm_{1-x}$Y$\rm_x$Cu$\rm_2$O$\rm_{8+\delta}$}

\def\y65cabi2212{Bi$\rm_2$Sr$\rm_2$Ca$\rm_{0.35}$Y$\rm_{0.65}$Cu$\rm_2$O$\rm_{8+\delta}$}

\def\Co{CeCoIn$_5$}
\def\Rh{CeRhIn$_5$}

\begin{document}

\title{Pressure Study of Quantum Criticality in CeCoIn$_5$}

\author{F.~Ronning,$^1$ C.~Capan,$^{1\ast}$ E.D. Bauer,$^1$ J.~D.~Thompson,$^1$ J.~L.~Sarrao$^1$, and R.~Movshovich,$^1$}
\affiliation{$^1$Los Alamos National Laboratory, Los Alamos, New
Mexico 87545}

\date{\today}

\begin{abstract}
We report resistivity measurements in the normal state of \Co~down
to 40 mK and simultaneously in magnetic fields up to 9 T in the
[001] crystallographic direction and under pressures up to 1.3
GPa. At ambient pressure the data are consistent with a field
tuned quantum critical point coincident with the superconducting
upper critical field $H_{c2}$, as observed previously. We find
that with increasing pressure the quantum critical point moves
inside the superconducting dome to lower fields. Thus, we can rule
out that superconductivity is directly responsible for the
non-Fermi liquid behavior in \Co. Instead, the data point toward
an antiferromagnetic quantum critical point scenario.
\end{abstract}
\maketitle

A quantum critical point is simply the point at which a second
order phase transition occurs at $T$ = 0, where quantum
fluctuations are present. Classical phase transitions are now well
understood. While theoretically there is a natural extension to
$T$ = 0,\cite{Hertz:PRB-76,Millis:PRB-92,Moriya:JPSJ-95} the
experimental systems (and in particular heavy fermion systems)
display serious discrepancies with these
predictions.\cite{Stewart:RMP-01} As disorder may profoundly
influence the behavior at a quantum critical point, there is a
great benefit in examining quantum critical systems which are
stoichiometric, and hence, relatively disorder free. \Co~is one of
a relatively small number of such systems.

\Co~is a heavy fermion superconductor with $T_c$~=~2.3
K.\cite{Petrovic:JPCM-01} The normal state possesses non-Fermi
liquid properties in zero field ($T$ linear resistivity, $T\ln(T)$
specific heat, and modified Curie-Weiss $\chi$, compared to the
Fermi liquid expectations of $T^2$ resistivity, $T$ linear
specific heat, and constant $\chi$) indicative of a nearby
underlying quantum critical point.\cite{Sidorov:prl-02,Kim:prb-01}
By applying the magnetic field along the tetragonal c-axis a
field-tuned quantum critical point (QCP) was identified at
$H_{QCP}$ = 5 T.\cite{JP1:prl-03,AndreaQCP:prl-03} The fact that
the superconducting upper critical field $H_{c2}$ is also at 5 T
raises the question of whether superconducting fluctuations could
be responsible for the field-tuned non-Fermi liquid behavior.
However, this observation (that $H_{c2} \approx H_{QCP}$) is
likely to be an accidental coincidence for several reasons: i) it
is not clear if a superconductor has sufficiently strong
fluctuations to produce an extended critical regime, ii) the
superconducting transition itself becomes first order below 0.7~K
in \Co,\cite{Andrea:prl-02} which should cutoff any diverging
fluctuations, and iii) similarities in the zero field
pressure-temperature phase diagrams of \Rh~and \Co~suggests that
\Co~at ambient pressure is in close proximity to an
antiferromagnetic quantum critical point, as is observed in
\Rh.\cite{Sidorov:prl-02} However, two experiments designed to
separate $H_{QCP}$ from $H_{c2}$, via magnetic field
anisotropy\cite{Ronning:prb-05} or Sn doping
studies\cite{Eric:prl-05}, failed to do so. Applying the magnetic
field in the ab-plane increases $H_{c2}$ to 12 T, while in Sn
doping studies the c-axis $H_{c2}$ was suppressed to as low as
2.75 T for CeCoIn$_{4.88}$Sn$_{0.12}$. Despite this variation in
$H_{c2}$ by more than a factor of 4, one did not observe the
appearance of an additional ordered phase above $H_{c2}$, nor a
Fermi-liquid regime at the superconducting upper critical field,
with either specific heat or resistivity measurements. This
suggests that the two critical fields are inherently linked
together.

The pressure phase diagram of \Rh~alluded to above suggests that
pressure may be an effective means of suppressing criticality in
\Co~if an antiferromagnetic QCP is indeed still the origin of the
non-Fermi liquid behavior in both systems. This is supported by
measurements of resistivity in zero field,\cite{Sidorov:prl-02}
specific heat,\cite{Sparn:PhysicaB-02} and
NQR\cite{Kohori:JMMM-04} which appear to restore Fermi liquid
behavior with increasing pressure in \Co. In addition, dHvA
results at high fields show the effective mass of the 2-D
cylindrical $\beta$ sheet (which increases as $H_{c2}$ is
approached from above) decreases with increasing
pressure.\cite{Shishido:JPCM-03} The evolution of the field-tuned
QCP with pressure is best identified by performing measurements
with magnetic field. This is precisely how $H_{QCP}$ was
originally identified to be close to $H_{c2}$ at ambient
pressure.\cite{JP1:prl-03,AndreaQCP:prl-03} In this paper we
report resistivity measurements of \Co~under pressure with
magnetic field up to 9 T applied along the c-axis and temperatures
down to 40 mK. We find that the QCP is strongly suppressed inside
the superconducting dome and hence is no longer coincident with
$H_{c2}$ as pressure is increased. This gives compelling evidence
that the origin of the non-Fermi liquid behavior is not associated
with superconductivity, but rather related to a order competing
with superconductivity, most likely antiferromagnetism.

Resistivity under pressure was obtained by a 4 point measurement
in a Cu-Be piston cell attached to the temperature regulation
stage on a dilution refrigerator. The use of Silicone as the
pressure transmitting medium ensures hydrostatic pressure
($\Delta$$P$ $\sim$ 0.01 GPa). A single crystal, aligned to within
5$^{\circ}$ of the c-axis, was measured in fields up to 9 T. Field
sweeps at 100 mK identified $H_{c2}$ as 4.95, 4.95, 4.63, and 4.0
T at the four measured pressures of 0, 0.05, 0.6, and 1.3 GPa. The
pressure was determined independently by measuring the
superconducting transition of a Sn sample by $ac$ susceptibility.

At ambient pressure the quantum critical field was determined by
tracking the divergence of the $T^2$ coefficient of resistivity in
the ever shrinking window of the Fermi liquid
regime.\cite{JP1:prl-03} We follow the same recipe here for four
different pressures: 0, 0.05, 0.6, and 1.3 GPa. The raw
resistivity data is shown versus $T^2$ in figure 1. Panels (a)
through (d) show data for various fields at a given pressure,
while panels (e) and (f) present resistivity curves for various
pressures at constant magnetic field. Recall that in the Fermi
liquid regime $\rho = \rho_0 + AT^2$. Thus, a straight line in
figure 1 would correspond to a well established Fermi liquid
behavior whose slope is proportional to the square of the
effective mass. At high fields (eg. 9 T; see fig. 1(f)) the Fermi
liquid regime extends over a wide temperature range. At low
temperatures there is an upturn in the data, which is believed to
originate from quasiparticles reaching the $\omega_c\tau \approx
1$ limit, and hence this is also a feature of the Fermi liquid
regime in \Co.\cite{JP1:prl-03} We note that a similar observation
was made in a very clean UPt$_3$
crystal.\cite{Taillefer:PhysicaC-88} For the purpose of analysis,
this prevents us from fitting $\rho = \rho_0 + AT^2$ down to the
lowest temperature measured; rather, we fit over a range which
maximizes the $A$ coefficient. As the magnetic field is reduced
toward $H_{c2}$ the temperature range of the $T^2$ Fermi liquid
regime decreases monotonically.

\begin{figure}
\includegraphics[width=3.3in]{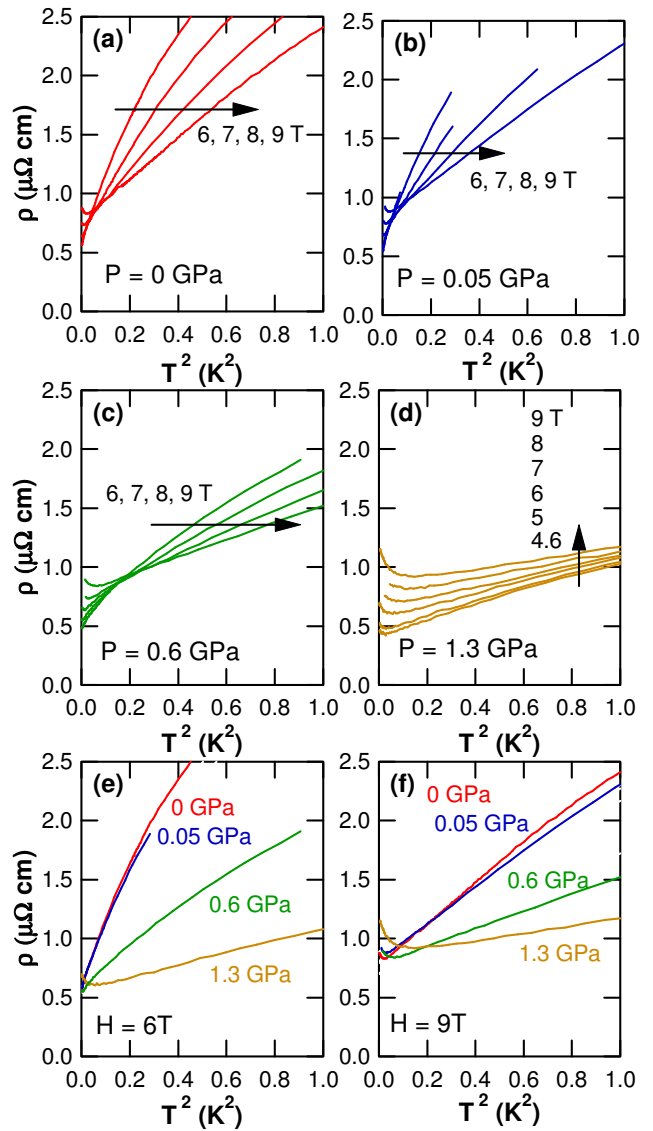}
\caption{\label{Resistivity-fig1} (color online) Resistivity of
\Co~ for $H \parallel$ c-axis and current in the ab-plane plotted
versus $T^2$ to highlight the Fermi liquid regime for fields above
the superconducting upper critical field. $H_{c2}$ = 4.95, 4.95,
4.63, and 4.0 T for P = 0, 0.05, 0.6, and 1.3 GPa, respectively.
Panels (a) through (d) are taken at the constant pressure
indicated, and panels (e) and (f) are four different pressures
measured at the same value of the magnetic field indicated in the
plot.}
\end{figure}

There are several features evident in the raw data which we will
attempt to quantify below. Beginning with the ambient pressure
data in fig. 1(a), there is a dramatic increase in the inelastic
scattering as the magnetic field is lowered. Simultaneously, the
$A$ coefficient grows, and the Fermi liquid regime becomes
vanishingly small. Let us contrast this with the 1.3 GPa data in
fig. 1(d). While the $A$ coefficient also increases with
decreasing field the effect is not nearly as dramatic as at
ambient pressure. Furthermore, there is still a well established
Fermi liquid at 4.6 T (which is just above $H_{c2}$ = 4 T) on a
temperature range significantly larger than at 6 T for ambient
pressure. This is shown both by a large linear regime on the plot
of $\rho$ versus $T^2$ and by the upturn still present at the
lowest measured temperatures. As for the resistivity data for 0.05
GPa and 0.6 GPa, we find it displays a smooth evolution between
the ambient pressure and 1.3 GPa extremes we just discussed. We
also should note that as the pressure increases the $A$
coefficient is also rapidly suppressed as can be seen in figures
1(e) and (f).

To quantify the above behavior in the Fermi liquid regime, we plot
the $A$ coefficient versus field in figure 2(a) for each measured
pressure. The divergence of the $A$ coefficient, and hence the
effective mass, is clearly suppressed with increasing pressure.
Since pressure is known to increase the Kondo temperature in Ce
Kondo lattice systems, we need to be certain that the reduction of
$A$ is not solely a decrease of the overall scale. For that
purpose, we parameterize the divergence of $A$ with the form $A =
A_0 (H - H_{QCP})^{\alpha}$ with $A_0$ and $H_{QCP}$ as adjustable
parameters. This form was used by Paglione {\it et al.} to
identify the critical field as $H_{QCP}$ = 5.1 T with $\alpha$ =
-1.37.\cite{JP1:prl-03} By keeping $\alpha$ fixed at -1.37 we find
both $A_0$ and $H_{QCP}$ to decrease with increasing pressure. The
values of $A_0$ are 14.1, 15.9, 12.0, and 8.2 $\mu\Omega$cm/K$^2$
for 0, 0.05, 0.6, and 1.3 GPa, respectively. The behavior of $A_0$
can be understood from the general view of increasing the Kondo
temperature and stabilizing the Fermi liquid with increasing
pressure, as commonly observed in Ce
compounds.\cite{Thompson:handbook-94} We attribute the decreasing
$H_{QCP}$ to the fact that the critical field is moving inside the
superconducting dome. The latter point is emphasized in figure
2(b). The critical fields from the fits above are plotted together
with $H_{c2}$. The fact that the critical field moves inside the
superconducting dome implies that we do not have a superconducting
critical point. It is interesting to note that the pressure at
which the extrapolated critical field reaches 0 T ($\sim$1.1 GPa)
is close to the pressure at which the superconducting transition
temperature is maximum ($\sim$1.3 GPa), a common feature of a wide
variety of quantum critical systems.\cite{Mathur:nature-98}

\begin{figure}
\includegraphics[width=3.3in]{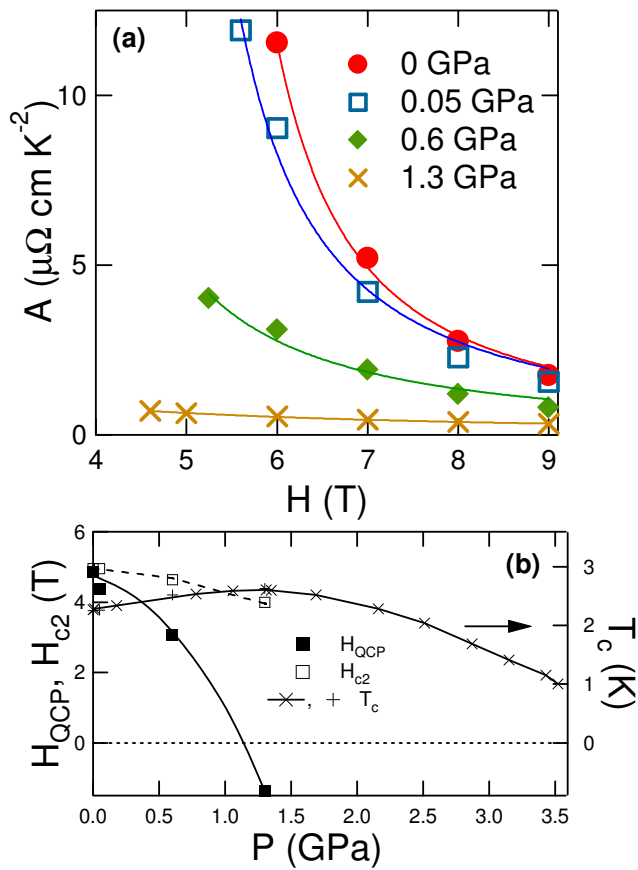}
\caption{\label{Resistivity-fig2} (color online) (a) The $T^2$
coefficient of the resistivity extracted from the data presented
in figure 1. The solid lines are fits to $A = A_0 (H -
H_{QCP})^{-1.37}$. The values of $H_{QCP}$ from the fits are
plotted in (b) along with the values of the superconducting upper
critical field and the superconducting $T_c$ in zero field versus
pressure. The $\times$'s are obtained from ref.
\cite{Sidorov:prl-02}. Lines in (b) are guides to the eye.}
\end{figure}

Finally, we extend our analysis of the data beyond the Fermi
liquid regime. Figure 3 presents an image plot of
$\partial\ln(\rho - \rho_0)/\partial\ln(T)$ in the $H-T$ plane for
$P$=0 and $P$=1.3 GPa. Similar plots have been made for
YbRh$_2$Si$_2$\cite{Custers:nature-03} and
Sr$_3$Ru$_2$O$_7$\cite{Grigera:Science-01} to help identify the
critical field. The logarithmic derivative  gives the value of the
exponent $n$ assuming the resistivity has the form $\rho = \rho_0
+ AT^n$. The low temperature regime of the resistivity upturn has
been suppressed, as the form of $\rho$ is inappropriate in this
regime. While for $P$=0 the data again suggests that the quantum
critical field occurs at the superconducting upper critical field
of 5 T, it does not seem possible to make a similar statement for
the P=1.3 GPa data. The open symbols mark the temperature where
the Fermi liquid fits deviate by more than 2$\%$ from the data.
This quantity should go to zero at the quantum critical point. At
ambient pressure we see that this Fermi crossover temperature is
rapidly vanishing as $H_{c2}$ is approached. At 1.3 GPa a linear
extrapolation yields a quantum critical point of 0.5 T, which is
well inside the superconducting upper critical field of 4 T, and
in reasonable agreement with our fit of the $A$ coefficient versus
magnetic field.

\begin{figure}
\includegraphics[width=3.3in]{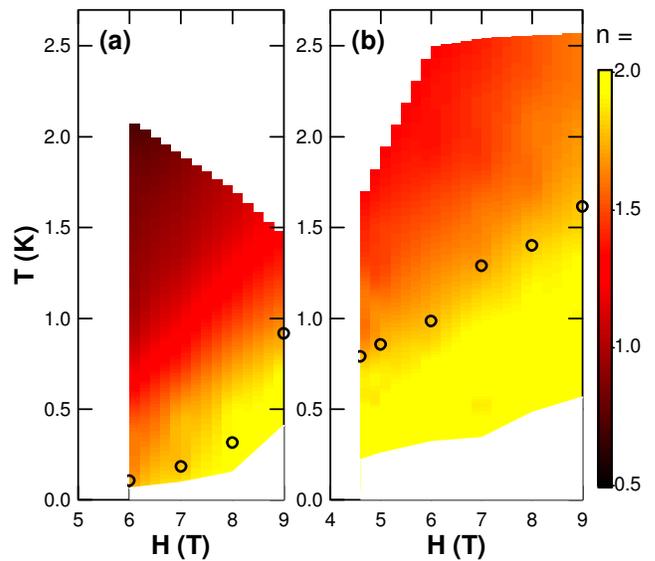}
\caption{\label{Resistivity-fig3} (color online) An image plot of
$\partial\ln(\rho - \rho_0)/\partial\ln(T)$ for (a) ambient
pressure and (b) 1.3 GPa. The open circles represent the Fermi
crossover temperature as discussed in the text.}
\end{figure}

Thus, by applying hydrostatic pressure we have now successfully
separated the critical field from the superconducting upper
critical field, and we conclude that it was merely an accidental
coincidence that $H_{QCP}$ = $H_{c2}$ at ambient pressure for
magnetic field along the c-axis.\cite{JP1:prl-03,AndreaQCP:prl-03}
Note that by adding an additional tuning parameter, namely
pressure, the field tuned critical point near $H_{c2}$, now
becomes a line of critical points in the $H-P$ plane. A
consequence of this is that, while ambient pressure \Co~possesses
a field tuned QCP, pressure tuning at zero field should find a
critical pressure which can be accessed with positive pressure. It
is hoped that measurements within the vortex cores will provide
more direct evidence of this.

One can next speculate as to the true origin of the quantum
critical fluctuations which produce the non-Fermi liquid behavior
in \Co. From the initial comparison of the \Rh~and \Co~phase
diagrams under pressure, it has been suggested that the origin of
the non-Fermi liquid behavior is the presence of an
antiferromagnetic quantum critical point.\cite{Sidorov:prl-02}
This speculation has been supported indirectly by numerous means.
Since one expects both magnetic field and pressure to suppress
antiferromagnetism in this compound, the fact that we observe the
quantum critical point to rapidly move within the superconducting
dome with increasing pressure is also consistent with this
picture. Recently, dHvA results on \Rh~under pressure show that
there is a quantum critical pressure of 2.35 GPa at which the
Fermi surface undergoes a local to itinerant crossover of the 4$f$
electrons.\cite{Shishido:JPSJ-05} Given that the high pressure
dHvA frequencies agree well with those observed in \Co~at ambient
pressure, it is tempting to suggest that \Co~at ambient pressure
lies slightly above this high field critical pressure.

While this provides a consistent perspective, there are still
several unresolved issues. There is no explanation for why Sn
doping was unable to separate the quantum critical point from the
superconducting upper critical field over such a wide range in
doping.\cite{Eric:prl-05} It is possible that Sn doping creates a
very extended non-Fermi liquid regime.\cite{Ramos-05} Furthermore,
if at ambient pressure in \Co~the critical field of 5~T refers to
a field suppression of an antiferromagnetic state, one must ask
the question as to why the magnetic transition has not been
observed at smaller fields? Perhaps this is tied to a typically
overlooked point that it is difficult to unify the non Fermi
liquid behavior at zero field with the slightly different
non-Fermi liquid properties observed at $H_{c2}$, and it is
possible that multiple critical points must be envoked to explain
all the features of \Co.\cite{JPCoQCP04} Our results also raise
several questions in the related compound, \Rh. It is necessary to
confirm in \Rh~that the critical pressure identified by
dHvA\cite{Shishido:JPSJ-05} does indeed correspond to an
antiferromagnetic quantum critical point. We would also like to
know if \Rh~has a similar line of critical points in the $H-P$
plane as identified here. We hope that future measurements can
resolve these issues.

In conclusion, we have studied resistivity of \Co~under pressure
and in high magnetic fields. We demonstrate that the quantum
critical field moves inside the superconducting dome with
increasing pressure, which rules out a novel superconducting
quantum critical point. This further suggests that the quantum
critical behavior is most likely associated with an as yet
undetected antiferromagnetic quantum critical point.

\begin{acknowledgments}
We would like to thank V. Sidorov and T. Park for helpful
discussions, the $T_c$ versus $P$ data shown in figure 2(b), and
assistance with the pressure cell. F.R. and E.D.B. acknowledge
support from the Reines Postdoctoral Fellowship (DOE-LANL). Work
at Los Alamos National Laboratory was performed under the auspices
of the U.S. Department of Energy.
\end{acknowledgments}


\begin{thebibliography}{25}
\expandafter\ifx\csname
natexlab\endcsname\relax\def\natexlab#1{#1}\fi
\expandafter\ifx\csname bibnamefont\endcsname\relax
  \def\bibnamefont#1{#1}\fi
\expandafter\ifx\csname bibfnamefont\endcsname\relax
  \def\bibfnamefont#1{#1}\fi
\expandafter\ifx\csname citenamefont\endcsname\relax
  \def\citenamefont#1{#1}\fi
\expandafter\ifx\csname url\endcsname\relax
  \def\url#1{\texttt{#1}}\fi
\expandafter\ifx\csname
urlprefix\endcsname\relax\def\urlprefix{URL }\fi
\providecommand{\bibinfo}[2]{#2}
\providecommand{\eprint}[2][]{\url{#2}}

\bibitem[$^\ast$]{Cigdem}Present address: Department of Physics, Louisianna State
University, Baton Rouge, LA 70803

\bibitem[{\citenamefont{JP1}(2003)}]{JP1:prl-03}
\bibinfo{author}{\bibfnamefont{J.} \bibnamefont{Paglione}},
\bibinfo{author}{\bibfnamefont{M.A.} \bibnamefont{Tanatar}},
\bibinfo{author}{\bibfnamefont{D.G.} \bibnamefont{Hawthorn}},
\bibinfo{author}{\bibfnamefont{E.} \bibnamefont{Boaknin}},
\bibinfo{author}{\bibfnamefont{R.W.} \bibnamefont{Hill}},
\bibinfo{author}{\bibfnamefont{F.} \bibnamefont{Ronning}},
\bibinfo{author}{\bibfnamefont{M.} \bibnamefont{Sutherland}},
\bibinfo{author}{\bibfnamefont{L.} \bibnamefont{Taillefer}},
\bibinfo{author}{\bibfnamefont{C.} \bibnamefont{Petrovic}},
\bibinfo{author}{\bibfnamefont{P.C.} \bibnamefont{Canfield}},
  \bibinfo{journal}{Phys. Rev. Lett.} \textbf{\bibinfo{volume}{91}},
  \bibinfo{pages}{246405} (\bibinfo{year}{2003}).

\bibitem[{\citenamefont{AndreaQCP}(2003)}]{AndreaQCP:prl-03}
\bibinfo{author}{\bibfnamefont{A.} \bibnamefont{Bianchi}},
\bibinfo{author}{\bibfnamefont{R.} \bibnamefont{Movshovich}},
\bibinfo{author}{\bibfnamefont{I.} \bibnamefont{Vekhter}},
\bibinfo{author}{\bibfnamefont{P.G.} \bibnamefont{Pagliuso}},
\bibinfo{author}{\bibfnamefont{J.L.} \bibnamefont{Sarrao}},
  \bibinfo{journal}{Phys. Rev. Lett.} \textbf{\bibinfo{volume}{91}},
  \bibinfo{pages}{257001} (\bibinfo{year}{2003}).

\bibitem[{\citenamefont{Hertz}(1976)}]{Hertz:PRB-76}
\bibinfo{author}{\bibfnamefont{J.A.} \bibnamefont{Hertz}},
  \bibinfo{journal}{Phys. Rev. B} \textbf{\bibinfo{volume}{14}},
  \bibinfo{pages}{1165} (\bibinfo{year}{1976}).

\bibitem[{\citenamefont{Millis}(1992)}]{Millis:PRB-92}
\bibinfo{author}{\bibfnamefont{A.J.} \bibnamefont{Millis}},
  \bibinfo{journal}{Phys. Rev. B} \textbf{\bibinfo{volume}{48}},
  \bibinfo{pages}{7183} (\bibinfo{year}{1993}).

\bibitem[{\citenamefont{Moriya}(2001)}]{Moriya:JPSJ-95}
\bibinfo{author}{\bibfnamefont{T.} \bibnamefont{Moriya}},
\bibinfo{author}{\bibfnamefont{T.} \bibnamefont{Takimoto}},
  \bibinfo{journal}{J. Phys. Soc. Japan} \textbf{\bibinfo{volume}{64}},
  \bibinfo{pages}{960} (\bibinfo{year}{1995}).


\bibitem[{\citenamefont{Stewart}(2001)}]{Stewart:RMP-01}
\bibinfo{author}{\bibfnamefont{G.R.} \bibnamefont{Stewart}},
  \bibinfo{journal}{Rev. Mod. Phys.} \textbf{\bibinfo{volume}{73}},
  \bibinfo{pages}{797} (\bibinfo{year}{2001}).

\bibitem[{\citenamefont{Petrovic}(2001)}]{Petrovic:JPCM-01}
\bibinfo{author}{\bibfnamefont{C.} \bibnamefont{Petrovic}},
\bibinfo{author}{\bibfnamefont{P.G.} \bibnamefont{Pagliuso}},
\bibinfo{author}{\bibfnamefont{M.F.} \bibnamefont{Hundley}},
\bibinfo{author}{\bibfnamefont{R.} \bibnamefont{Movshovich}},
\bibinfo{author}{\bibfnamefont{J.L.} \bibnamefont{Sarrao}},
\bibinfo{author}{\bibfnamefont{J.D.} \bibnamefont{Thompson}},
\bibinfo{author}{\bibfnamefont{Z.} \bibnamefont{Fisk}},
\bibinfo{author}{\bibfnamefont{P.} \bibnamefont{Monthoux}},
  \bibinfo{journal}{J. Phys. Cond. Mat.} \textbf{\bibinfo{volume}{13}},
  \bibinfo{pages}{L337} (\bibinfo{year}{2001}).

%
\bibitem[{\citenamefont{Kim}(2001)}]{Kim:prb-01}
\bibinfo{author}{\bibfnamefont{J.S.} \bibnamefont{Kim}},
\bibinfo{author}{\bibfnamefont{J.} \bibnamefont{Alwood}},
\bibinfo{author}{\bibfnamefont{G.R.} \bibnamefont{Stewart}},
\bibinfo{author}{\bibfnamefont{J.L.} \bibnamefont{Sarrao}},
\bibinfo{author}{\bibfnamefont{J.D.} \bibnamefont{Thompson}},
  \bibinfo{journal}{Phys. Rev. B} \textbf{\bibinfo{volume}{64}},
  \bibinfo{pages}{134524} (\bibinfo{year}{2001}).

  \bibitem[{\citenamefont{Sidorov}(2002)}]{Sidorov:prl-02}
\bibinfo{author}{\bibfnamefont{V.A.} \bibnamefont{Sidorov}},
\bibinfo{author}{\bibfnamefont{M.} \bibnamefont{Nicklas}},
\bibinfo{author}{\bibfnamefont{P.G.} \bibnamefont{Pagliuso}},
\bibinfo{author}{\bibfnamefont{J.L.} \bibnamefont{Sarrao}},
\bibinfo{author}{\bibfnamefont{Y.} \bibnamefont{Bang}},
\bibinfo{author}{\bibfnamefont{A.V.} \bibnamefont{Balatsky}},
\bibinfo{author}{\bibfnamefont{J.D.} \bibnamefont{Thompson}},
  \bibinfo{journal}{Phys. Rev. Lett.} \textbf{\bibinfo{volume}{89}},
  \bibinfo{pages}{157004} (\bibinfo{year}{2002}).

\bibitem[{\citenamefont{Andrea}(2002)}]{Andrea:prl-02}
\bibinfo{author}{\bibfnamefont{A.} \bibnamefont{Bianchi}},
\bibinfo{author}{\bibfnamefont{R.} \bibnamefont{Movshovich}},
\bibinfo{author}{\bibfnamefont{N.} \bibnamefont{Oeschler}},
\bibinfo{author}{\bibfnamefont{P.} \bibnamefont{Gegenwart}},
\bibinfo{author}{\bibfnamefont{F.} \bibnamefont{Steglich}},
\bibinfo{author}{\bibfnamefont{J.D.} \bibnamefont{Thompson}},
\bibinfo{author}{\bibfnamefont{P.G.} \bibnamefont{Pagliuso}},
\bibinfo{author}{\bibfnamefont{J.L.} \bibnamefont{Sarrao}},
  \bibinfo{journal}{Phys. Rev. Lett.} \textbf{\bibinfo{volume}{89}},
  \bibinfo{pages}{137002} (\bibinfo{year}{2002}).

\bibitem[{\citenamefont{Eric}(2004)}]{Ronning:prb-05}
\bibinfo{author}{\bibfnamefont{F.} \bibnamefont{Ronning}},
\bibinfo{author}{\bibfnamefont{C.} \bibnamefont{Capan}},
\bibinfo{author}{\bibfnamefont{A.} \bibnamefont{Bianchi}},
\bibinfo{author}{\bibfnamefont{R.} \bibnamefont{Movshovich}},
\bibinfo{author}{\bibfnamefont{A.} \bibnamefont{Lacerda}},
\bibinfo{author}{\bibfnamefont{M.F.} \bibnamefont{Hundley}},
\bibinfo{author}{\bibfnamefont{J.D.} \bibnamefont{Thompson}},
\bibinfo{author}{\bibfnamefont{P.G.} \bibnamefont{Pagliuso}},
\bibinfo{author}{\bibfnamefont{J.L.} \bibnamefont{Sarrao}},
  \bibinfo{journal}{Phys. Rev. B} \textbf{\bibinfo{volume}{71}},
  \bibinfo{pages}{104528} (\bibinfo{year}{2005}).

\bibitem[{\citenamefont{Eric}(2004)}]{Eric:prl-05}
\bibinfo{author}{\bibfnamefont{E.D.} \bibnamefont{Bauer}},
\bibinfo{author}{\bibfnamefont{C.} \bibnamefont{Capan}},
\bibinfo{author}{\bibfnamefont{F.} \bibnamefont{Ronning}},
\bibinfo{author}{\bibfnamefont{R.} \bibnamefont{Movshovich}},
\bibinfo{author}{\bibfnamefont{J.D.} \bibnamefont{Thompson}},
\bibinfo{author}{\bibfnamefont{J.L.} \bibnamefont{Sarrao}},
  \bibinfo{journal}{Phys. Rev. Lett.} \textbf{\bibinfo{volume}{94}},
  \bibinfo{pages}{047001} (\bibinfo{year}{2005}).

\bibitem[{\citenamefont{Sparn}(2002)}]{Sparn:PhysicaB-02}
\bibinfo{author}{\bibfnamefont{G.} \bibnamefont{Sparn}},
\bibinfo{author}{\bibfnamefont{R.} \bibnamefont{Borth}},
\bibinfo{author}{\bibfnamefont{E.} \bibnamefont{Lengyel}},
\bibinfo{author}{\bibfnamefont{P.G.} \bibnamefont{Pagliuso}},
\bibinfo{author}{\bibfnamefont{J.L.} \bibnamefont{Sarrao}},
\bibinfo{author}{\bibfnamefont{F.} \bibnamefont{Steglich}},
\bibinfo{author}{\bibfnamefont{J.D.} \bibnamefont{Thompson}},
  \bibinfo{journal}{Physica B} \textbf{\bibinfo{volume}{319}},
  \bibinfo{pages}{262} (\bibinfo{year}{2002}).

\bibitem[{\citenamefont{Kohori}(2004)}]{Kohori:JMMM-04}
\bibinfo{author}{\bibfnamefont{Y.} \bibnamefont{Kohori}},
\bibinfo{author}{\bibfnamefont{H.} \bibnamefont{Saito}},
\bibinfo{author}{\bibfnamefont{Y.} \bibnamefont{Kobayashi}},
\bibinfo{author}{\bibfnamefont{H.} \bibnamefont{Taira}},
\bibinfo{author}{\bibfnamefont{Y.} \bibnamefont{Iwamoto}},
\bibinfo{author}{\bibfnamefont{T.} \bibnamefont{Kohara}},
\bibinfo{author}{\bibfnamefont{T.} \bibnamefont{Matsumoto}},
\bibinfo{author}{\bibfnamefont{E.D.} \bibnamefont{Bauer}},
\bibinfo{author}{\bibfnamefont{M.B.} \bibnamefont{Maple}},
\bibinfo{author}{\bibfnamefont{J.L.} \bibnamefont{Sarrao}},
  \bibinfo{journal}{J. Mag. Mag. Mat.} \textbf{\bibinfo{volume}{272-276}},
  \bibinfo{pages}{189} (\bibinfo{year}{2004}).

\bibitem[{\citenamefont{Shishido}(2003)}]{Shishido:JPCM-03}
\bibinfo{author}{\bibfnamefont{H.} \bibnamefont{Shishido}},
\bibinfo{author}{\bibfnamefont{T.} \bibnamefont{Ueda}},
\bibinfo{author}{\bibfnamefont{S.} \bibnamefont{Hashimoto}},
\bibinfo{author}{\bibfnamefont{T.} \bibnamefont{Kubo}},
\bibinfo{author}{\bibfnamefont{R.} \bibnamefont{Settai}},
\bibinfo{author}{\bibfnamefont{H.} \bibnamefont{Harima}},
\bibinfo{author}{\bibfnamefont{Y.} \bibnamefont{Onuki}},
  \bibinfo{journal}{J. Phys. Cond. Mat.} \textbf{\bibinfo{volume}{15}},
  \bibinfo{pages}{L499} (\bibinfo{year}{2003}).

\bibitem[{\citenamefont{Taillefer}(2002)}]{Taillefer:PhysicaC-88}
\bibinfo{author}{\bibfnamefont{L.} \bibnamefont{Taillefer}},
\bibinfo{author}{\bibfnamefont{F.} \bibnamefont{Piquemal}},
\bibinfo{author}{\bibfnamefont{J.} \bibnamefont{Flouquet}},
  \bibinfo{journal}{Physica C} \textbf{\bibinfo{volume}{153}},
  \bibinfo{pages}{451} (\bibinfo{year}{1988}).


\bibitem[{\citenamefont{Shishido}(2003)}]{Shishido:JPSJ-05}
\bibinfo{author}{\bibfnamefont{H.} \bibnamefont{Shishido}},
\bibinfo{author}{\bibfnamefont{R.} \bibnamefont{Settai}},
\bibinfo{author}{\bibfnamefont{H.} \bibnamefont{Harima}},
\bibinfo{author}{\bibfnamefont{Y.} \bibnamefont{Onuki}},
  \bibinfo{journal}{J. Phys. Soc. Japan} \textbf{\bibinfo{volume}{74}},
  \bibinfo{pages}{1103} (\bibinfo{year}{2005}).

%
\bibitem[{\citenamefont{Mathur}(1998)}]{Mathur:nature-98}
\bibinfo{author}{\bibfnamefont{N.D.} \bibnamefont{Mathur}},
\bibinfo{author}{\bibfnamefont{F.M.} \bibnamefont{Grosche}},
\bibinfo{author}{\bibfnamefont{S.R.} \bibnamefont{Julian}},
\bibinfo{author}{\bibfnamefont{I.R.} \bibnamefont{Walker}},
\bibinfo{author}{\bibfnamefont{D.M.} \bibnamefont{Freye}},
\bibinfo{author}{\bibfnamefont{R.K.W.} \bibnamefont{Haselwimmer}},
\bibinfo{author}{\bibfnamefont{G.G.} \bibnamefont{Lonzarich}},
  \bibinfo{journal}{Nature} \textbf{\bibinfo{volume}{394}},
  \bibinfo{pages}{39} (\bibinfo{year}{1998}).

%
\bibitem[{\citenamefont{Thompson}(1998)}]{Thompson:handbook-94}
\bibinfo{author}{\bibfnamefont{J.D.} \bibnamefont{Thompson}},
\bibinfo{author}{\bibfnamefont{J.M.} \bibnamefont{Lawrence}},
  \bibinfo{journal}{Handbook of the Physics and Chemistry of the Rare Earths}
  \textbf{\bibinfo{volume}{19}},
  \bibinfo{pages}{383} (\bibinfo{year}{1994}).

\bibitem[{\citenamefont{Custers}(2003)}]{Custers:nature-03}
\bibinfo{author}{\bibfnamefont{J.} \bibnamefont{Custers}},
\bibinfo{author}{\bibfnamefont{P.} \bibnamefont{Gegenwart}},
\bibinfo{author}{\bibfnamefont{H.} \bibnamefont{Wilhelm}},
\bibinfo{author}{\bibfnamefont{K.} \bibnamefont{Neumaier}},
\bibinfo{author}{\bibfnamefont{Y.} \bibnamefont{Tokiwa}},
\bibinfo{author}{\bibfnamefont{O.} \bibnamefont{Trovarelli}},
\bibinfo{author}{\bibfnamefont{C.} \bibnamefont{Geibel}},
\bibinfo{author}{\bibfnamefont{F.} \bibnamefont{Steglich}},
\bibinfo{author}{\bibfnamefont{C.} \bibnamefont{Pepin}},
\bibinfo{author}{\bibfnamefont{P.} \bibnamefont{Coleman}},
  \bibinfo{journal}{Nature} \textbf{\bibinfo{volume}{424}},
  \bibinfo{pages}{524} (\bibinfo{year}{2003}).


\bibitem[{\citenamefont{Grigera}(2001)}]{Grigera:Science-01}
\bibinfo{author}{\bibfnamefont{S.A.} \bibnamefont{Grigera}},
\bibinfo{author}{\bibfnamefont{R.S.} \bibnamefont{Perry}},
\bibinfo{author}{\bibfnamefont{A.J.} \bibnamefont{Schofield}},
\bibinfo{author}{\bibfnamefont{M.} \bibnamefont{Chiao}},
\bibinfo{author}{\bibfnamefont{S.R.} \bibnamefont{Julian}},
\bibinfo{author}{\bibfnamefont{G.G.} \bibnamefont{Lonzarich}},
\bibinfo{author}{\bibfnamefont{S.I.} \bibnamefont{Ikeda}},
\bibinfo{author}{\bibfnamefont{Y.} \bibnamefont{Maeno}},
\bibinfo{author}{\bibfnamefont{A.J.} \bibnamefont{Millis}},
\bibinfo{author}{\bibfnamefont{A.P.} \bibnamefont{Mackenzie}},
  \bibinfo{journal}{Science} \textbf{\bibinfo{volume}{294}},
  \bibinfo{pages}{329} (\bibinfo{year}{2001}).

\bibitem[{\citenamefont{Ramos}(2005)}]{Ramos-05}
\bibinfo{author}{\bibfnamefont{S.M.} \bibnamefont{Ramos}},
\bibinfo{author}{\bibfnamefont{M.B.} \bibnamefont{Fontes}},
\bibinfo{author}{\bibfnamefont{A.D.} \bibnamefont{Alverenga}},
\bibinfo{author}{\bibfnamefont{E.} \bibnamefont{Baggio-Saitovitch}},
\bibinfo{author}{\bibfnamefont{P.G.} \bibnamefont{Pagliuso}},
\bibinfo{author}{\bibfnamefont{E.D.} \bibnamefont{Bauer}},
\bibinfo{author}{\bibfnamefont{J.D.} \bibnamefont{Thompson}},
\bibinfo{author}{\bibfnamefont{J.L.} \bibnamefont{Sarrao}},
\bibinfo{author}{\bibfnamefont{M.A.} \bibnamefont{Continentino}},
  \bibinfo{journal}{Physica B} \textbf{\bibinfo{volume}{359}},
  \bibinfo{pages}{398} (\bibinfo{year}{2005}).

\bibitem[{\citenamefont{JPCoQCP}(2004)}]{JPCoQCP04}
\bibinfo{author}{\bibfnamefont{J.} \bibnamefont{Paglione}},
\bibinfo{author}{\bibfnamefont{M.A.} \bibnamefont{Tanatar}},
\bibinfo{author}{\bibfnamefont{D.G.} \bibnamefont{Hawthorn}},
\bibinfo{author}{\bibfnamefont{E.} \bibnamefont{Boaknin}},
\bibinfo{author}{\bibfnamefont{F.} \bibnamefont{Ronning}},
\bibinfo{author}{\bibfnamefont{R.W.} \bibnamefont{Hill}},
\bibinfo{author}{\bibfnamefont{M.} \bibnamefont{Sutherland}},
\bibinfo{author}{\bibfnamefont{L.} \bibnamefont{Taillefer}},
\bibinfo{author}{\bibfnamefont{C.} \bibnamefont{Petrovic}},
\bibinfo{author}{\bibfnamefont{P.C.} \bibnamefont{Canfield}},
  \bibinfo{journal}{cond-mat} \textbf{\bibinfo{volume}{}},
  \bibinfo{pages}{0405157} (\bibinfo{year}{2004}).


\end{thebibliography}
\end{document}